\documentclass[11pt]{article}
\usepackage{amssymb,amsmath,cite,geometry,graphicx,url}
\catcode`\@=11

\catcode`\@=11
\def\thesection{\arabic{section}.}

\def\appendix{\setcounter{section}{0}
        \def\thesection{Appendix.}
        \def\theequation{\Alph{section}.\arabic{equation}}}
\def\section{\@startsection{section}{1}{\z@}{3.5ex plus 1ex minus
   .2ex}{2.3ex plus .2ex}{\large\bf}}
     
\long\def\@makefntext#1{\parindent 0cm\noindent
\hbox to 1em{\hss$^{\@thefnmark}$}#1}

\newcommand{\captionfonts}{\small}
\makeatletter  
\long\def\@makecaption#1#2{%
  \vskip\abovecaptionskip
  \sbox\@tempboxa{{\captionfonts #1: #2}}%
  \ifdim \wd\@tempboxa >\hsize
    {\captionfonts #1: #2\par}
  \else
    \hbox to\hsize{\hfil\box\@tempboxa\hfil}%
  \fi
  \vskip\belowcaptionskip}
\makeatother   

\begin{document}
\begin{titlepage}
\vspace{.5in}
\begin{flushright}
November 2019\\  
\end{flushright}
\vspace{.5in}
\begin{center}
{\Large\bf
A comment on\\ 
``How the cosmological constant is hidden by\\[.6ex]
 Planck scale curvature fluctuations''}\\  
\vspace{.4in}
{S.~C{\sc arlip}\footnote{\it email: carlip@physics.ucdavis.edu}\\
       {\small\it Department of Physics}\\
       {\small\it University of California}\\
       {\small\it Davis, CA 95616}\\{\small\it USA}}
\end{center}

\vspace{.5in}
\begin{center}
{\large\bf Abstract}
\end{center}
\begin{center}
\begin{minipage}{4.2in}
{\small
A recent preprint by Wang and Unruh [arXiv:1911.06110] contains a number
of criticisms of my paper, ``Hiding the cosmological constant'' 
[Phys.\ Rev.\ Lett.\ 123 (2019) 131302, arXiv:1809.08277].  While Wang 
and Unruh suggest an interesting alternative scenario and raise an important 
conceptual question, most of their criticisms are incorrect, in part because 
of misunderstandings about averaging and about the nature of the ``foamy'' 
spacetimes considered in my paper.}
\end{minipage}
\end{center}
\end{titlepage}
\addtocounter{footnote}{-1}

In Ref.\ \cite{Carlip}, I proposed a new approach to the cosmological constant problem, 
suggesting that perhaps the Universe really \emph{does} have a large cosmological
constant, but one that is effectively hidden by Planck scale fluctuations of
geometry and topology, Wheeler's ``spacetime foam'' \cite{Wheeler}.  Since we
observe the Universe at lengths much larger than the Planck length, such a
scenario necessarily involves averaging.  As a simple example, I considered volume 
averaging on a spatial slice,
\begin{equation}
\langle X\rangle_{\mathcal{U}} = \frac{1}{V_{\mathcal{U}}}\int_{\mathcal{U}}  X\sqrt{g}\,d^3x 
   \quad\text{with} 
   \quad V_{\mathcal{U}} = \int_{\mathcal{U}} \sqrt{g}\,d^3x \, ,
\label{x1}
\end{equation}
where the region $\mathcal{U}$ is defined in some time-independent way.  I argued
that for a very large set of ``foamy'' initial data, the average $\langle K\rangle$ of
the trace of the extrinsic curvature was zero, and that a time-slicing could be
chosen for which $\langle K\rangle$ remained zero, at least for short times, on
constant time hypersurfaces.

As part of the analysis, I used the fact that
\begin{equation}
\frac{d\ }{dt}\langle K\rangle 
   =  \frac{1}{V_{\mathcal{U}}}\int_{\mathcal{U}} N\left(\Lambda + \frac{2}{3}K^2 
   - 2\sigma^2\right)\sqrt{g}\,d^3x \, ,
\label{x2}
\end{equation}
where $\sigma^2 = \frac{1}{2}\sigma^{ij}\sigma_{ij}$ is the square of the shear.%
\footnote{This is not, of course, new; it appeared, for instance, in \cite{Buchert} twenty
years ago.}  Since the sign of the integrand in (\ref{x2}) is indefinite, there exists a 
choice of positive lapse function for which the integral vanishes.  I argued that the 
same was true for higher derivatives, implying the existence, at least in a neighborhood of the
initial slice, of a foliation in which $\langle K\rangle = 0$ on each time slice.

In version 1 of Ref.\ \cite{Wang}, Wang and Unruh mistakenly claimed that eqn.\
(\ref{x2}) was wrong.  They have corrected this error, but continue to argue the time
derivatives in (\ref{x2}) should be changed to proper time derivatives 
$$\frac{d\ }{d\tau} = \frac{1}{N}\frac{d\ }{dt} \, .$$
If one were interested in the behavior at a single point, this might make sense.  Averaged, 
such a calculation would perhaps tell us about the rate of change of the expansion as 
seen by a particular choice of ``average observer.''  
These are interesting questions, but they are not the questions discussed in 
Ref.\ \cite{Carlip}.  Rather, Ref.\ \cite{Carlip} asked the following: if $\langle K\rangle=0$  
on an initial slice, is there any foliation in which $\langle K\rangle$ remains zero on 
future slices?  This is a question about derivatives of an average, not the average of 
derivatives.  Differentiation and averaging do not commute,
\begin{equation}
\frac{1}{N}\frac{d\langle X\rangle}{dt} \ne \left\langle  \frac{1}{N}\frac{dX}{dt} \right\rangle \, ,
\label{x5}
\end{equation}
and if $N$ varies rapidly, the difference can be quite large.   

Ref.\ \cite{Wang} claims that this question---does there exist a foliation?---is not physical,
because ``a choice of lapse corresponds to a choice of coordinates, and no physics can 
depend purely on the choice of coordinates.''  This is, of course, incorrect.  When we ask
whether the Universe is homogeneous, we are asking about the existence of a foliation
that has certain properties.  If we start with a homogeneous spatial slice, future slices 
will only be homogeneous with certain choices of the lapse.  If we ask whether the Hubble 
constant is really spatially constant, this is again a question about the existence of a 
foliation; indeed, it is almost the same as the question asked in Ref.\ \cite{Carlip}.  Of
course it would be nice to have a manifestly coordinate-independent phrasing of such
questions.  But regardless of the phrasing, the existence or nonexistence of foliations
with certain properties is a physical statement.

Wang and Unruh further contend that Planck scale fluctuations cannot hide a positive cosmological 
constant.  To do so, they appeal to the geodesic deviation equation, which they 
say shows that ``arbitrary free falling observers and arbitrary nearby free falling test particles''
necessarily diverge exponentially.  Their argument relies on a linear approximation, however: 
while the geodesic deviation equation is exact, both the inertial frame used in \cite{Wang} and 
the interpretation of the deviation vector as a distance are only valid infinitesimally.   One can 
learn more by looking at the Raychaudhuri equation, which described the behavior of a 
whole ``pencil''  of geodesics.  For a congruence of timelike geodesics, this takes the
form \cite{Wald}
\begin{equation}
\frac{d\theta}{d\tau} = -\frac{1}{3}\theta^2 - \sigma_{ab}\sigma^{ab} 
    + \omega_{ab}\omega^{ab} + \Lambda \, ,
\label{xa1}
\end{equation}
where $\theta$, $\sigma_{ab}$, and $\omega_{ab}$ are the expansion, shear, and
vorticity of the congruence.  Here the expansion has a clear physical meaning: it
is the fractional rate of change of volume, with respect to proper time, of the region
filled by the geodesics.  If $\theta=0$, the pencil of geodesics can twist and shear, 
but the volume it comprises remains constant.  

Just as in (\ref{x2}), a positive cosmological constant causes geodesics to diverge,
but the divergence can be counteracted by shear.  This is a kind of local version of 
the results of \cite{Waldb}, where it was shown that a cosmological constant in a 
homogeneous Bianchi IX universe tends to lead to exponential expansion, but
only if $\Lambda$ is large compared to spatial curvature terms.  In a somewhat
different context, the shear term in (\ref{xa1}) can be interpreted physically as a 
gravitational wave energy density \cite{Ashtekar}; the Raychaudhuri equation 
may be read as a statement that the contraction caused by such energy
can compensate for the expansion caused by $\Lambda$.

Note in particular that the spacetimes considered in \cite{Carlip} contain large
numbers of marginally outer trapped surfaces (MOTS) \cite{Burkhartb}.  Indeed, if 
the ``foamy'' features occur at the Planck scale, such MOTS occur at Planck density.  
The resulting trapped regions are essentially interiors of black holes,\footnote{Technical 
results regarding black holes require additional assumptions about the behavior 
of null infinity that may not hold here, but the qualitative features are the same.}  in 
which even null geodesics necessarily converge, so geodesic balls certainly do not
expand exponentially.  

There are, in fact, a few rigorous results in the literature on the asymptotic 
behavior of inhomogeneous vacuum spacetimes with a positive cosmological constant
\cite{Kleban,Mirbabayi,Moncrief}.  But these all require a Cauchy surface that either 
has constant mean curvature ($K = \hbox{const.}$) or is ``everywhere expanding''  
($K>0$ everywhere).  The spacetimes considered in \cite{Carlip} do not satisfy the first 
condition, and it seems highly unlikely that they satisfy the second.  The question of
the generality of asymptotically de Sitter behavior is certainly an important one, but
an answer will take a good deal more than an approximate local solution of the geodesic
deviation equation.
 
Having cleared the underbrush, we can now turn to the substantive issue raised
in Ref.\ \cite{Wang}:  whether $\langle K\rangle=0$ necessarily implies that 
the cosmological constant is hidden.  Using the example of the static patch of de Sitter
space, Wang and Unruh argue that it does not.  This is an important but difficult subject,
intertwined with the notorious ``problem of time'' and questions how to define averaged
observers and to convert coordinate-dependent results into invariant statements.

To start with, Wang and Unruh propose a criterion that 
\begin{equation}
\left\langle\frac{D_iN}{N}\right\rangle = 0 \, ,
\label{x6}
\end{equation}
which, they argue, means that the relevant observers at constant $x^i$ are not, on the 
average, accelerating.  For a single geodesic with vanishing initial spatial velocity and
a fixed time-slicing, this is indeed the condition that the spatial velocity remain zero
\cite{Gour}.  There are, as always, delicacies in averaging---if a fluctuation makes the
velocity momentarily nonzero at some point, the effect is nonlinear---but (\ref{x6}) is at
least a reasonable first-order condition.  

Although it was not discussed explicitly in Ref.\ \cite{Carlip}, the lapse functions considered 
there almost certainly obey this condition.  At the initial surface, the lapse $N$ is to be
chosen to make the right-hand side of (\ref{x2}) vanish.  But by construction, the initial 
data are chosen in such a way that $K^2$ and $\sigma^2$ have rapid and essentially
random spatial variations.  There is no way for such a structure to have a preferred
direction; if in one Planck-sized region $D_iN$ points in a particular direction, there
will be a nearby region in which it points in the opposite direction.

This is not yet the whole story, though.  The basic construction in \cite{Carlip} depends on
an arbitrary initial slice, and even with the requirement $\langle K\rangle=0$,  condition
(\ref{x6}) is not enough to pick out a unique lapse function.  It may be that the full answer 
requires a more fundamentally quantum treatment---it was speculated in \cite{Carlip} that 
the ``foamy'' structure of the initial slice might reproduce itself in time---but we can try to 
go further  classically.  

The basic problem is a long-standing one: how to choose a time-slicing that has a clear
physical significance, so that statements about time dependence have an unambiguous 
meaning.  For a spacetime with a Killing or conformal Killing vector, that vector determines 
a choice of time.  For a homogeneous spacetime, the hypersurfaces of homogeneity 
define a time-slicing.  But for a highly inhomogeneous spacetime with no symmetries, 
the problem is much harder.

In mathematical relativity, there are two standard choices of time-slicing: maximal
slicing, in which $K$ (often written as ``$\mathrm{Tr}\,K$'') is zero, and York time,
$K = -t$.  For a static spacetime, Killing time gives a maximal slicing; for an FLRW 
cosmology, York time gives the same foliation as the standard cosmological time.  For
the spacetimes considered in \cite{Carlip}, neither of these choices is available:
the construction typically excludes even a single hypersurface of constant $K$.
The proposal of \cite{Carlip} is essentially to use the next best thing, averaged 
maximal slicing, $\langle K\rangle = 0$.  If, as suggested in \cite{Wangb}, higher
order calculations give a small residual cosmological constant, perhaps we will be
able to define averaged York time.

Ultimately, though, we need to connect the predictions of these models to honest physical
observables.  Were it not for the need to average, this might not be too hard.  For
instance, for a null geodesic with affine parameter $\lambda$, wave vector $k^\mu$, 
and frequency $k^0=\omega$,  it may be checked that 
\begin{equation}
\frac{d\ }{d\lambda}(N\omega) = -(k^i\partial_iN)\omega + K_{ij}k^ik^j \, ,
\label{x7}
\end{equation}
so if $\langle D_iN\rangle$ and $\langle K_{ij}\rangle$ both vanish, the proper
frequency receives no red shift.  The problem, though, is that we do not observe
red shifts at the Planck scale; we need to analyze quantities like average frequencies
for light waves with wavelengths much larger than the inhomogeneities in the
geometry.  Similarly, we might try to model physical clocks to see what a choice of 
their synchronization means, but this again requires averaging over regions much
larger than the inhomogeneities.   

None of this means that the specific proposal in Ref.\ \cite{Wang} for hiding a
negative cosmological constant is necessarily wrong.  It does make it clear, though,
that a more careful treatment of averaging is needed.  A related analysis, albeit in
a different context, has been carried out in \cite{Buchert}, and a result quite similar 
to eqn.\ (24) of Ref.\ \cite{Wang} appears there, but with closer attention to
averaging.  Perhaps this might be useful.  In addition, adjacent Planck-scale oscillating 
regions of \cite{Wang} must somehow be joined together to form a single manifold.  
The only way I know of to ``glue'' such regions without introducing domain walls is 
to employ the construction of Chrusciel et al.\ \cite{Chrusciel,Chruscielb} used in Ref.\ 
\cite{Carlip}.   But as  noted above, this gluing leads to additional trapped regions and 
singularities \cite{Burkhartb}, which may compete with the cosmological singularities 
that play an essential role in the proposal of \cite{Wang}.  Again, more careful analysis 
is needed.

\end{document}